\def\acbar{{\sc Acbar\ }}

\documentclass[11pt,preprint2]{aastex}

\usepackage{epsfig}
\usepackage{changebar}
\usepackage{amsmath}

\bibpunct{(}{)}{;}{a}{}{,}

\slugcomment{To be published in the proceedings of "The Cosmic 
Microwave Background and its Polarization", New 
Astronomy Reviews, (eds. S. Hanany and K.A. Olive).}

\shorttitle{First Results from \sc{Acbar}}
\shortauthors{Runyan et al.}

\usepackage{amsmath}
\usepackage{amssymb}
\begin{document}

\title{First Results from the Arcminute Cosmology Bolometer Array Receiver}

\author{\small M.C. Runyan\altaffilmark{1,2,3},
P.A.R. Ade\altaffilmark{4},
J.J. Bock\altaffilmark{5}, 
J.R. Bond\altaffilmark{6},
C. Cantalupo\altaffilmark{7},
C.R. Contaldi\altaffilmark{6},
M.D. Daub\altaffilmark{8}, 
J.H. Goldstein\altaffilmark{9,10}, 
P.L. Gomez\altaffilmark{11},
W.L. Holzapfel\altaffilmark{8},
C.L. Kuo\altaffilmark{8,12}, 
A.E. Lange\altaffilmark{1},
M. Lueker\altaffilmark{8},
M. Newcomb\altaffilmark{8},
J.B. Peterson\altaffilmark{11},
D. Pogosyan\altaffilmark{6}, 
A.K. Romer\altaffilmark{11},
J. Ruhl\altaffilmark{9}, 
E. Torbet\altaffilmark{10},
D. Woolsey\altaffilmark{8}}

\altaffiltext{1}{Department of Physics, Math, and Astronomy, California Institute 
of Technology, Pasadena, CA 91125}
\altaffiltext{2}{Current address: University of Chicago, LASR-132, 5640 S. Ellis Ave.,
Chicago, IL 60637, mcr@grizelda.uchicago.edu}
\altaffiltext{3}{Enrico Fermi Fellow, CfCP Fellow}
\altaffiltext{4}{Department of Physics and Astronomy, Cardiff University, CF24 3YB Wales, UK}
\altaffiltext{5}{Jet Propulsion Laboratory, Pasadena, CA 91125}
\altaffiltext{6}{Canadian Institute for Theoretical Astrophysics, University of Toronto, Canada}
\altaffiltext{7}{Lawrence Berkeley National Laboratory, Berkeley, CA 94720}
\altaffiltext{8}{Department of Physics, University of California at Berkeley, Berkeley, CA 94720}
\altaffiltext{9}{Department of Physics, Case Western Reserve University, Cleveland, OH 44106}
\altaffiltext{10}{Department of Physics, University of California, Santa Barbara, CA 93106}
\altaffiltext{11}{Department of Physics, Carnegie Mellon University, Pittsburgh, PA 15213}
\altaffiltext{12}{Department of Astronomy, University of California at Berkeley, Berkeley, CA 94720}

\begin{abstract}
{\small We review the first science results from the 
Arcminute Cosmology Bolometer Array Receiver 
({\sc Acbar}); a multifrequency millimeter-wave receiver optimized for observations 
of the Cosmic Microwave Background (CMB) and the Sunyaev-Zel'dovich (SZ) effect in 
clusters of galaxies. {\sc Acbar} was installed on the 2 m Viper telescope at the 
South Pole in January 2001 and the results presented here incorporate data through
July 2002.  We present the power spectrum of the CMB at 150 GHz 
over the range $\ell=150 - 3000$ measured 
by {\sc Acbar} as well as estimates for the values of the cosmological parameters 
within the context of $\Lambda$CDM models.  We find that the inclusion of $\Omega_\Lambda$
greatly improves the fit to the power spectrum.  We also observe a slight 
excess of small-scale anisotropy at 150 GHz; if interpreted as power
from the SZ effect of unresolved clusters, the measured signal is consistent with
CBI and BIMA within the context of the SZ power spectrum models tested.
}
\end{abstract}


\section{Introduction}\label{sec:intro}

In this proceedings we review the first science results
from the \acbar experiment.
We present an overview of the receiver, telescope, and site in
\S\ref{sec:instrument}. Our observing technique is
described in \S\ref{sec:observations}.  The 
map making technique and power spectrum estimation
are described in \S\ref{sec:powerspec}.
We discuss systematic tests, including
a check on foreground contamination, in \S\ref{sec:sys}.  Estimation
of cosmological parameters is presented in \S\ref{sec:params} and
we discuss the results in \S\ref{sec:discussion}.  
Further details of the \acbar power spectrum
and cosmological parameter extraction are presented in \citet{kuo02} and
\citet{goldstein02}, respectively.  Due to journal
space constraints, we refer the reader to associated papers on \acbar pointed SZ 
cluster observations \citep{romer03} and blind cluster survey
\citep{runyan03b}.  


\section{Instrument and Telescope}\label{sec:instrument}

The Arcminute Cosmology Bolometer Array Receiver ({\sc Acbar}) is
a 16 pixel, millimeter wavelength, 240 mK bolometer array and is described
in detail in \citet{runyan03a}.  The instrument was designed to 
couple to the existing Viper telescope at the South Pole to produce
high resolution maps of the CMB sky with high signal-to-noise.


\acbar is configurable to observe simultaneously at 150, 220,
280 and 350 GHz with bandwidths of 31, 31, 48, and 24 GHz, respectively.
In 2001 \acbar had four feeds at each of the four observing frequencies.
In 2002 we replaced the 350 GHz feeds with an additional row
of 150 GHz feeds because of the
superior noise performance at 150 GHz.  
\acbar makes use of extremely sensitive microlithographed 
spider-web bolometers developed at JPL for the Planck satellite
mission \citep{turnerbolo}.  These detectors achieve background photon
limited performance in {\sc Acbar}.  In 2002 the 150 GHz channels
had an average $NET_{RJ}$ sensitivity of $\sim200~\mu{\rm K}\sqrt{s}$.
The focal plane is arranged in a $4\times 4$ grid with a spacing of
$\sim 16^\prime$ between beam centers on the sky.  

The Viper telescope is a 2 m off-axis Gregorian telescope designed
specifically for observations of CMB anisotropy.  The primary is surrounded
by a 0.5 m skirt to reflect spillover to the sky.  The
entire telescope is enclosed in a large conical ground shield to 
block emission from elevations below $\sim 25^\circ$; one section
lowers to allow observations of low-elevation sources such as
planets.  A chopping flat at the image of the primary
formed by the secondary sweeps the beams $\sim 3^\circ$ in azimuth in 
a fraction of a second.

The combination of large chop ($\sim 3^\circ$) and small beam sizes
($\sim 4-5'$ FWHM Gaussians) 
makes {\sc Acbar} sensitive to a wide range of angular
scales ($150<\ell<3000$), with high $\ell$-space resolution ($\Delta
\ell\sim 150$). Another unique feature of {\sc Acbar} is its multi-frequency
coverage, which has the potential to discriminate between
sources of signal and foreground confusion.
The CMB power spectrum \citep{kuo02} and cosmological constraints \citep{goldstein02}
presented here are derived from the
$150\,$GHz channel data collected from January 2001 through July 2002. 
Analysis of the $220\,$GHz and $280\,$GHz data and the remainder of
the 2002 150 GHz data is underway.

The South Pole provides an exceptional platform from which to conduct
millimeter-wave observations.  The elevation at the Pole is $\sim{9,300}^\prime$
and the average ambient temperature during the austral winter is near $-80^\circ$F.
The high altitude, dry air, and lack of diurnal variation result in a 
transparent and extremely stable atmosphere \citep{lay00,peterson02}.  
The entire
southern celestial hemisphere is available year-round allowing 
very deep integrations.  When combined with a well established research
infrastructure, these attributes makes the 
South Pole an ideal location for terrestrial CMB observations.

\section{CMB Observations}\label{sec:observations}

To minimize possible pickup from the modulation of telescope sidelobes
on the ground shield, 
we restrict the CMB observations to fields with ${\rm EL} \gtrsim 45^\circ$. 
The power spectrum reported in \citet{kuo02} is derived from observations of
two separate fields, which we call CMB2 and CMB5. 
The remaining two CMB fields from 2002 are currently being analyzed.

High signal-to-noise maps of planets (Mars in July 2001,
and Venus in September 2002) are used to accurately measure the beam 
patterns of the array elements
as well as calibrate the instrument.
We have estimated the total calibration
uncertainty to be 10\% and further details of the instrument calibration
can be found in \citet{runyan03a}.

Each CMB field was selected to include a bright, flat-spectrum radio
source.
The coadded image of the guide source is used to determine the effective 
beam sizes at the center of the map and includes smearing due
to pointing jitter that occurs over the period in which the data are acquired.
These final point-source image sizes 
are consistent with the beam sizes measured on planets and 
the observed pointing RMS determined from frequent observations of galactic 
sources.  

When the telescope is chopping the time-stream signals are dominated by
a quadratic chopper-synchronous signal roughly 10 mK in amplitude
at 150 GHz.  The chopper
signal is due mostly to snow accumulation on the telescope.  We employ an
observing strategy that allows us to remove both constant and linearly
time-varying offsets by observing three adjacent 
fields (LEAD, MAIN, and TRAIL) in succession.  
By forming the difference $LMT = M - (L+T)/2$ we remove the chopper offsets
while preserving large-scale CMB power on the sky.

\section{Power Spectrum}\label{sec:powerspec}

Algorithms for the analysis of total power CMB data are well developed and 
have been tested on balloon-borne experiments \citep{netterfield02, lee01}. 
However, the ground-based {\sc Acbar} experiment is subject to constraints 
which require significant departures from the standard analysis
algorithms. Here we briefly outline the {\sc Acbar} analysis and highlight
its unique features.  Complete details of the analysis is presented in \citet{kuo02}.


In \citet{kuo02} we developed a {\it cleaned noise-weighted coadded
map} as an intermediate step from time stream data to power spectrum.
This technique identifies periods when the data have significant
correlated atmospheric noise and adaptively projects out the 
corrupted spatial modes.
The data are also weighted by their variance after projection of the
corrupted modes.  The cleaned and weighted map is,
however, not an unbiased
estimator of the CMB sky temperature; the projection
of modes and noise weighting must be accounted for in the
theory and noise covariance matrices.

\begin{figure}
\centering
\epsfig{figure=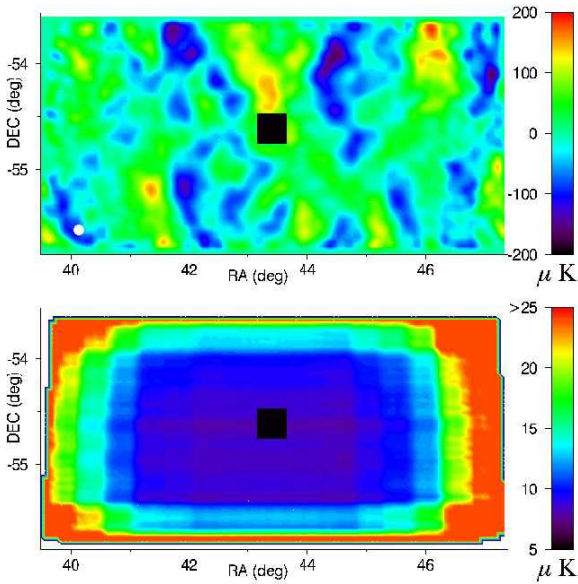,width=3in,clip=}
\caption{\protect\small
The top panel shows the LMT-differenced, atmospheric mode removed,
noise weighted, coadded map for the CMB5 field.
The guide quasar has been replaced with black pixels.
The small white circle in the lower left hand corner of the map
represents the FWHM of the average array element beam size as determined from
the coadded quasar image.  The map is pixelated at $2.5^\prime$ and has
been smoothed with a $4.5^\prime$ FWHM Gaussian.
The predominance of extended structure in the vertical direction results from
de-projection of extended horizontal structure during atmospheric mode removal.
The lower panel shows the noise in the LMT differenced map as a function of
position.  The S/N the degree-scale structures in this map approach 100. 
This figure is from \citet{kuo02}.}
\label{fig:cmb5_map}
\end{figure}

The LMT-differenced map
of the CMB5 field is shown in Figure \ref{fig:cmb5_map}. 
Due to computational limitations, we choose the map pixelization to be
$2.5^{\prime}$.  
For the purposes of presentation, we have smoothed the pixelized map
with a $4.5^{\prime}$ FWHM Gaussian beam. 
The lower panel in Figure \ref{fig:cmb5_map} shows the 
RMS noise in the LMT difference map as a function of position.
Due to differences in sky coverage, the noise varies across the map;
in the central region, the RMS noise per $5'$ beam is found to be
$17\,\mu$K and $8\,\mu$K for the CMB2 and CMB5 LMT-differenced maps, respectively. 
On degree angular scales, the S/N in the center of the CMB5
map approaches 100.  
These maps have not yet had the undetected PMN source catalog and IRAS dust
templates projected out.
Foregrounds are discussed in more detail in \S\ref{sec:sys}.

\begin{figure}
\centering
\epsfig{file=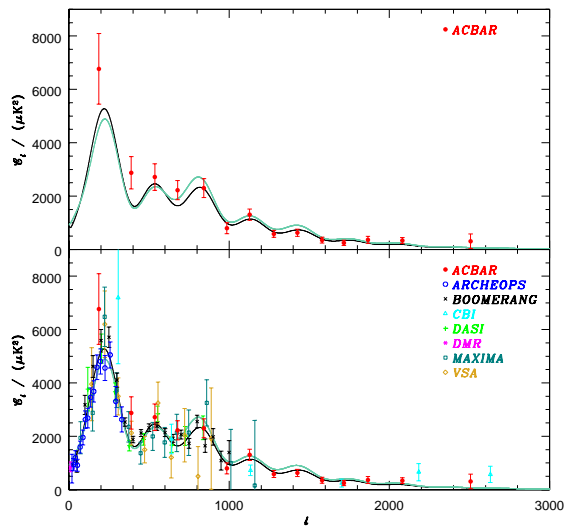,width=3.0in,clip=}
\caption{\protect\small
Top Panel:  The \acbar CMB power spectrum, ${\mathcal C}_{\ell} \equiv \ell
(\ell + 1)C_{\ell}/{(2\pi)}$, plotted over a vacuum energy
dominated [$\Omega_k = -0.05$, $\Omega_\Lambda = 0.5$, $\omega_{cdm} = 0.12$,
$\omega_b = 0.02$, $H_0 = 50$, $\tau_C = 0.025$, $n_s = 0.925$, amplitude
${\mathcal C}_{10}= 1.11\times 10^{-10}T_{\rm CMB}^2$]
model (black thin line)
and a CDM dominated [$\Omega_k = 0.05$, $\Omega_\Lambda = 0$, 
$\omega_{cdm} = 0.22$, $\omega_b = 0.02$, $H_0 = 50$, $\tau_C = 0$,
$n_s = 0.925$, amplitude 
${\mathcal C}_{10}= 1.34\times 10^{-10}T_{\rm CMB}^2$] model
(green thick line).
These are the best-fit models,
for $\Lambda$ and $\Lambda$-free models respectively,
found during the {\sc Acbar}+Others parameter estimation described 
\citet{goldstein02},
with the weak-$h$ prior.
Bottom Panel: The top panel with the addition of power
spectra from several other experiments.  
Both models appear to be reasonable fits to the data, with the 
$\Omega_\Lambda = 0.5$ model statistically being the better of the two.
This figure is from \citet{goldstein02}.}
\label{fig:aps}  
\end{figure}

The maximum likelihood band powers are estimated iteratively
using the quadratic iteration method \citep{bond98}.
The \acbar power spectrum -- as well as the power spectra from
other contemporary experiments -- is shown in Figure \ref{fig:aps}.
The decorrelated band powers and window functions are available
on the \acbar website\footnote{http://cosmology.berkeley.edu/group/swlh/acbar/}.

\section{Systematic Tests}\label{sec:sys}

We performed a number of systematic checks upon the \acbar data to verify the
robustness of the measured power spectrum.  Of particular concern
was the effect of possible foreground contamination.  We selected the target
regions on the sky to have low dust contrast from the IRAS/DIRBE maps
of \citet{finkbeiner99} extrapolated to 150 GHz.  
We estimate the total variance in the LMT-differenced maps
due to dust at 150 GHz to be about 9 $\mu$K$^2$ and 70 $\mu$K$^2$ for CMB2 
and CMB5, respectively.  In addition, unresolved point sources may also 
contribute power to the maps.  We project out the central pointing sources
in each field and a bright radio galaxy (Pictor A) in the CMB2 field.  We 
detect only one other PMN radio source at greater than $3\sigma$ and it too
is removed.  We investigated the potential 
effects of these foregrounds by generating the power
spectrum with no additional 
foreground removal and then re-calculating it after projecting
out the IRAS/DIRBE template and all 170 PMN sources in the CMB fields above 40
mJy at 5 GHz.  The foreground removal has a negligible impact on the power
spectrum indicating an insignificant level of contamination. 
We use the foreground-removed power spectrum for all subsequent analysis.

We also performed a number of ``jackknife'' tests on the data.  We divided each CMB
data set into a first-half and second-half and calculated the power spectra to
look for systematic changes over time.  We
also broke the data into left-going and right-going chopper sweeps to look
for direction-dependent effects.  Finally, we divided the data into $180^\circ$
azimuth chunks either towards or away from the near-by Martin A. Pomerantz Observatory
(MAPO) building to investigate pick-up from sources at low elevation.  The power
spectrum passed all three jackknife tests within the error bars (determined by
Monte Carlo simulation).

\section{Cosmological Parameter Extraction}\label{sec:params}

\subsection{Method}\label{sec:parammethod}

In this section we describe using the CMB power spectrum measured
by \acbar and other CMB experiments to derive
Bayesian estimates of cosmological parameters in inflation-motivated adiabatic
CDM models.  A more complete treatment is presented in \citet{goldstein02}.
Our set of cosmological parameters includes 
$\Omega_k=(1-\Omega_{tot})$, $\Omega_{\Lambda}$, 
$\omega_{cdm}$, $\omega_b$, $n_s$, $\tau_C$,
$\ln{{\mathcal C}_{10}}$.  The total energy density ($\Omega_{tot}$) is the
sum of the energy density from the vacuum ($\Omega_\Lambda$), matter, and
relativistic particles.  The matter density is split into two constituents:
baryonic matter ($\Omega_b \equiv \omega_b/h^2$) and cold dark matter
($\Omega_{cdm} \equiv \omega_{cdm}/h^2$), where $h$ is the Hubble parameter
in units of 100 km/s/Mpc.   $n_s$ is the scalar index of
primordial perturbations.
The amplitude of the power spectrum at
$\ell=10$ ($\ln{{\mathcal C}_{10}}$) gives the overall amplitude of the 
primordial fluctuations and has been well constrained by the COBE-DMR 
\citep{bennett96} [used in the parameter analysis in \citet{goldstein02}] 
and more recently by WMAP \citep{hinshaw03}. 

The universe reionized at some point between decoupling and the
present.  After reionization, CMB photons scatter further. $\tau_C$ is
the Compton optical depth (from decoupling to present) due to such
scattering.  High $\tau_C$ diminishes CMB power by a factor of
$\exp[-2\tau_C]$ over most of the $\ell$ range in the $TT$ spectrum, 
though not in the DMR
range.  After the parameter estimates in \citet{goldstein02}
were derived, the WMAP team reported a significant detection of 
the reionization signature with a best-fit $\tau_C = 0.17\pm 0.04$ 
from an excess in the $TE$ cross-spectrum at low-$\ell$
\citep{kogut03}.  The \acbar results have assumed no prior on
$\tau_C$ but re-analysis in the light of the WMAP results is
underway.

Because of the high resolution of {\sc Acbar}, it is possible that
sources of secondary anisotropy, such as from the Sunyaev-Zel'dovich
effect in clusters, could contribute significant power above the
primary CMB spectrum at high-$\ell$.  However,
the detection of power in the highest $\ell$ bin is only $0.9\sigma$
above the best-fit model CMB power spectrum.  Thus, for cosmological parameter
estimation based on the full \acbar power spectrum 
we believe we can safely ignore the effects from the potential
SZ contamination.  As the precision of our high-$\ell$ data improves,
however, we will have to include the SZ contribution in deriving
parameter estimates.


We also derive four observable quantities from the seven basic parameters.  These
are the present age of the Universe ($t_0$), the Hubble parameter ($h$), the
variance in the linear density fluctuation spectrum smoothed on 
$8h^{-1}$ Mpc scales ($\sigma_8^2$), and the effective shape parameter of the 
linear density power spectrum ($\Gamma_{eff}$).  The values of the derived parameters
are calculated at each point on the parameter grid but do not reflect
additional information;
they are used for incorporating prior constraints from other data sets,
such as the value of the Hubble parameter from HST or the value of 
$\sigma_8 \Omega_m^{0.56}$ from large scale structure (LSS) surveys.  The
individual priors are described in detail in \citet{goldstein02}.

\subsection{Results}\label{sec:paramresults}

By applying the method described above we can investigate the power
of adding \acbar to the various data sets.  We can also investigate
the effects of applying different priors upon the results.
As a first test we computed the 1D parameter likelihoods by adding
\acbar to the COBE-DMR power spectrum \citep{bond98}.  The thought
is that the COBE data provides a low-$\ell$ anchor to the power spectrum
and the precise high-$\ell$ measurements of \acbar will be sensitive to
the physics of the damping tail.  In particular, the strength of the
viscous and diffusive couplings are sensitive to baryon density, $\omega_b$
\citep{white01}.

However, what we found is that the degenerate influence of other cosmological
parameters upon the damping tail ({\it e.g.,} $\Omega_M$ and $n_s$)
prevents a precise measurement of $\omega_b$ without prior knowledge
from the peak-dip structure at lower-$\ell$ (degenerates are always 
causing problems such as this).  
In addition, some of the parameter results are found to depend upon
assumed priors.  Once we include information on the power spectrum from
experiments at lower $\ell$ the \acbar data improves constraints on parameters
by virtue of the small error bars.  The best-fit parameters of the joint
data set are found to be stable to the application of priors and are
reproduced in Table \ref{tbl:params}.

\begin{deluxetable}{llllllllllll}
\tablecolumns{12}
\tabletypesize{\tiny}
\tablecaption{Parameter Estimates and Errors}
\tablewidth{0pt}
\tablehead{
\colhead{Priors}
& \colhead{Run}
& \colhead{$\Omega_{tot}$}
& \colhead{$n_s$}
& \colhead{$\Omega_bh^2$}
& \colhead{$\Omega_{cdm}h^2$}
& \colhead{$\Omega_{\Lambda}$}
& \colhead{$\Omega_m$}
& \colhead{$\Omega_b$}
& \colhead{$h$}
& \colhead{Age}
& \colhead{$\tau_C$}
}
\startdata
\multicolumn{2}{l}{weak--$h$} \\
& Others
& $1.03^{0.05}_{0.04}$ 
& $0.96^{0.09}_{0.05}$ 
& $0.022^{0.003}_{0.002}$ 
& $0.13^{0.03}_{0.03}$ 
& $0.53^{0.18}_{0.19}$ 
& $0.50^{0.19}_{0.19}$ 
& $0.072^{0.023}_{0.023}$ 
& $0.57^{0.11}_{0.11}$ 
& $14.9^{1.3}_{1.3}$ 
& $<0.48$
\vspace{2pt}
\\
& \acbar+Others
& $1.04^{0.04}_{0.04}$ 
& $0.95^{0.09}_{0.05}$ 
& $0.022^{0.003}_{0.002}$ 
& $0.12^{0.03}_{0.03}$ 
& $0.57^{0.16}_{0.18}$ 
& $0.47^{0.18}_{0.18}$ 
& $0.071^{0.022}_{0.022}$ 
& $0.57^{0.11}_{0.11}$ 
& $15.1^{1.3}_{1.3}$ 
& $<0.47$ 
\\
\multicolumn{2}{l}{HST--$h$} \\
& Others
& $1.00^{0.03}_{0.03}$ 
& $0.99^{0.07}_{0.07}$ 
& $0.022^{0.003}_{0.003}$ 
& $0.12^{0.03}_{0.02}$ 
& $0.68^{0.09}_{0.12}$ 
& $0.33^{0.11}_{0.11}$ 
& $0.049^{0.013}_{0.013}$ 
& $0.68^{0.08}_{0.08}$ 
& $13.7^{1.0}_{1.0}$ 
& $<0.45$ 
\vspace{2pt}
\\
& \acbar+Others
& $1.00^{0.03}_{0.02}$ 
& $0.97^{0.07}_{0.06}$ 
& $0.022^{0.003}_{0.002}$ 
& $0.12^{0.02}_{0.02}$ 
& $0.70^{0.07}_{0.10}$ 
& $0.31^{0.10}_{0.10}$ 
& $0.049^{0.013}_{0.013}$ 
& $0.68^{0.08}_{0.08}$ 
& $13.9^{0.9}_{0.9}$ 
& $<0.43$ 
\\
\multicolumn{2}{l}{wk--$h$+flat} \\
& Others
& (1.00) 
& $0.95^{0.08}_{0.05}$ 
& $0.022^{0.003}_{0.002}$ 
& $0.13^{0.03}_{0.03}$ 
& $0.59^{0.15}_{0.23}$ 
& $0.43^{0.19}_{0.19}$ 
& $0.056^{0.014}_{0.014}$ 
& $0.63^{0.10}_{0.10}$ 
& $13.9^{0.5}_{0.5}$ 
& $<0.34$ 
\vspace{2pt}
\\
& \acbar+Others
& (1.00) 
& $0.95^{0.07}_{0.05}$ 
& $0.022^{0.002}_{0.002}$ 
& $0.13^{0.02}_{0.02}$ 
& $0.66^{0.10}_{0.16}$ 
& $0.35^{0.15}_{0.15}$ 
& $0.049^{0.011}_{0.011}$ 
& $0.67^{0.09}_{0.09}$ 
& $13.8^{0.4}_{0.4}$ 
& $<0.31$ 
\\
\\[0.01cm]\tableline\\[0.01cm]
$$
& \acbar+Others
\\
\multicolumn{2}{l}{wk--$h$+LSS}
$$
& $1.03^{0.05}_{0.04}$ 
& $0.98^{0.09}_{0.07}$ 
& $0.022^{0.003}_{0.003}$ 
& $0.11^{0.02}_{0.03}$ 
& $0.64^{0.08}_{0.12}$ 
& $0.41^{0.11}_{0.11}$ 
& $0.067^{0.019}_{0.019}$ 
& $0.59^{0.09}_{0.09}$ 
& $15.2^{1.4}_{1.4}$ 
& $<0.51$ 
\vspace{2pt}
\\
\multicolumn{2}{l}{wk--$h$+flat+LSS}
$$
& (1.00) 
& $0.94^{0.07}_{0.05}$ 
& $0.022^{0.002}_{0.002}$ 
& $0.13^{0.02}_{0.02}$ 
& $0.65^{0.08}_{0.11}$ 
& $0.36^{0.10}_{0.10}$ 
& $0.050^{0.008}_{0.008}$ 
& $0.66^{0.07}_{0.07}$ 
& $13.9^{0.4}_{0.4}$ 
& $<0.32$ 
\vspace{2pt}
\\
\multicolumn{2}{l}{wk--$h$+flat+LSS($low$--$\sigma_8$)}
$$
& (1.00) 
& $0.98^{0.07}_{0.06}$ 
& $0.022^{0.002}_{0.002}$ 
& $0.12^{0.02}_{0.02}$ 
& $0.71^{0.06}_{0.07}$ 
& $0.28^{0.07}_{0.07}$ 
& $0.044^{0.006}_{0.006}$ 
& $0.71^{0.07}_{0.07}$ 
& $13.7^{0.4}_{0.4}$ 
& $<0.34$ 
\vspace{2pt}
\\
\multicolumn{2}{l}{{\em strong data}}
$$
& $1.01^{0.03}_{0.02}$ 
& $0.99^{0.07}_{0.05}$ 
& $0.023^{0.003}_{0.002}$ 
& $0.12^{0.02}_{0.02}$ 
& $0.70^{0.05}_{0.05}$ 
& $0.31^{0.05}_{0.05}$ 
& $0.051^{0.011}_{0.011}$ 
& $0.67^{0.05}_{0.05}$ 
& $14.1^{0.9}_{0.9}$ 
& $<0.49$ 
\vspace{2pt}
\\
\multicolumn{2}{l}{{\em strong data}+flat}
$$
& (1.00) 
& $0.97^{0.05}_{0.05}$ 
& $0.022^{0.002}_{0.002}$ 
& $0.12^{0.01}_{0.01}$ 
& $0.70^{0.04}_{0.05}$ 
& $0.30^{0.04}_{0.04}$ 
& $0.046^{0.004}_{0.004}$ 
& $0.69^{0.04}_{0.04}$ 
& $13.7^{0.2}_{0.2}$ 
& $<0.32$ 
\vspace{2pt}
\\
\multicolumn{2}{l}{{\em strong data}+flat+LSS($low$--$\sigma_8$)}
$$
& (1.00) 
& $0.97^{0.05}_{0.05}$ 
& $0.022^{0.002}_{0.002}$ 
& $0.12^{0.01}_{0.01}$ 
& $0.71^{0.05}_{0.04}$ 
& $0.28^{0.05}_{0.05}$ 
& $0.045^{0.004}_{0.004}$ 
& $0.70^{0.04}_{0.04}$ 
& $13.7^{0.2}_{0.2}$ 
& $<0.31$ 
\\
\enddata
\tablecomments{\small
Parameter estimates and errors for several prior combinations
with and without {\sc Acbar}.  
Errors are quoted at 1--$\sigma$ (16\% and 84\%
points of the integral of the likelihood), except for $\tau_C$ where
the 95\% upper-limit is given.  The various priors are described in 
\citet{goldstein02}.  The top block lists results found with and without the
inclusion of \acbar data, which shows the small improvements found
upon adding \acbar to the mix.  The bottom block shows the effect
of applying stronger priors on the {\sc Acbar}+Others dataset, which 
naturally leads to improvements on the parameter estimates.
The difference between the LSS and LSS(low-$\sigma_8$) priors 
does lead to several slight shifts, smaller
than the 1--$\sigma$ errors.  This table is reproduced from \citet{goldstein02}. }
\label{tbl:params}
\end{deluxetable}

One would expect that the addition of high sensitivity data at high-$\ell$ would
lead to significant improvements upon the parameter that controls the scale 
of damping ($\omega_b$) and the tilt ($n_s$) because of the increased
$\ell$ baseline; but that appears not to be the case, as evidenced in
Table \ref{tbl:params}.  It is likely that the near degeneracies between certain
parameter pairs [see, for example, \citet{efstathiou99}]
implies that improved precision of measurement does not necessarily
translate into significant 
improvements in the 1D parameter likelihoods.  However, we can measure the
power of a data set to constrain cosmological parameter space by determining
the set of parameter ``eigenmodes'' that are most well constrained 
by the likelihood.  We calculated the parameter eigenmodes and associated eigenvalue
uncertainty for the set of CMB data before \acbar and then re-calculated the
eigenvectors including the {\sc Acbar} data.

Most of the eigenvectors are dominated by one or two parameters and it 
is interesting to note that the eigenvector most improved
by inclusion of the \acbar data is dominated by $\Omega_\Lambda$.
The eigenvectors are orthogonal and thus the product of 
their uncertainties represents a ``volume'' of parameter space consistent with the
data.  We find that including the \acbar data set reduces the volume of acceptable
$\Lambda$CDM parameter space by a factor of $\sim3$.  This shows that roughly
$2/3$ of the previously consistent parameter space has been eliminated by 
inclusion of the \acbar data, 
but this is not reflected in the marginalized likelihoods of the fiducial 
parameters.  

As mentioned above, the addition 
of the \acbar data has a substantial impact on the likelihood of
models without a cosmological constant.  
The $3\sigma$ lower-limit is increased from $\Omega_\Lambda > 0.086$ for the
``Other'' CMB experiments to $\Omega_\Lambda > 0.136$ by including {\sc Acbar}.  
The $\chi^2$ of best-fit free-$\Omega_\Lambda$ and $\Omega_\Lambda=0$ models is
$\chi^2 = 140$ and $\chi^2 = 160$ for 116 band powers, respectively.
The probably that the dark energy density is zero has become significantly 
unlikely.

In addition to estimating cosmological parameters from the primary power
spectrum, we also derived constraints on $\sigma_8$ from secondary anisotropy
resulting from a background of unresolved SZ clusters \citep{komatsu02}.
We employed two model SZ power spectra: one generated from Smoothed Particle
Hydrodynamics (SPH) simulations \citep{bond02b} and the other an analytical
model \citep{zhang02}.  The amplitude of the SZ power spectrum is expected to 
depend very strongly upon the value of $\sigma_8$ going roughly as 
${\mathcal C}_\ell^{SZ}\sim (\Omega_b h)^2\sigma_8^7$.  This strong dependence 
should allow significant constraints to be placed upon $\sigma_8$ with
existing (and to a higher degree with forthcoming) high-$\ell$ data sets from
{\sc Acbar}, CBI \citep{mason02}, and BIMA \citep{dawson02}.  

For this analysis we only use power spectrum data points above
$\ell\gtrsim{1500}$.  Although the primary power spectrum is
falling rapidly for $\ell{>}2000$, its contribution compared to the SZ signal
is by no means negligible; particularly at
150 GHz.  We include the contribution from the best-fit power 
spectrum of the ``{\sc Acbar}+Others'' data set in our analysis 
($\Omega_b=0.047$, $\Omega_{cdm}=0.253$, $\Omega_\Lambda=0.7$,
$h=0.69$, $n_s=0.975$, and $\tau_C = 0.2$).  We account for
the difference between our best-fit cosmology and the
one used to generate the SZ power spectra ($\Omega_b h = 0.035$) by
using $\sigma_8^{SZ} = (\Omega_b h/0.035)^{0.29}\sigma_8$.
We also include the effects of the non-Gaussian nature of the
SZ signal upon the sample variance [see \citet{goldstein02} for 
a discussion].

We find a best-fit value of $\sigma_8^{SZ}=0.98^{+0.12}_{-0.21}$ for the
joint {\sc Acbar}, CBI, and BIMA data set using the analytical
model and a slightly higher value of 1.04 for the SPH model.
These values of $\sigma_8$ are on the high-end of values
determined from LSS measurements.  
It is possible that the excess
power on small scales measured by CMB experiments may be the 
result of high-redshift supernova from the first stars \citep{oh03}.
On the other hand, it may simply reflect a current lack of understanding of
the SZ power spectrum.
The best-fit values of $\sigma_8^{SZ}$ measured by \acbar at 150 GHz
are consist with those measured by CBI and BIMA at 30 GHz within
$1\sigma$.  Although the \acbar data alone do not yield a $3\sigma$ lower
limit to $\sigma_8$, the combination of {\sc Acbar}, CBI, and BIMA
data results in a $3\sigma$ lower limit of $\sigma_8^{SZ}>0.63$ 
(including non-Gaussian effects) within the context of
the SZ models tested.

\section{Discussion}\label{sec:discussion}

The \acbar experiment has been used to precisely measure the CMB sky
from the South Pole 
and has produced the highest signal-to-noise map of the CMB to
date.  This data has yielded the most sensitive measurement of the 
damping tail region of the CMB power spectrum as of this writing;
analysis of the remainder of the 2001 through 2002 \acbar data is
underway and should further refine our power spectrum and parameter
estimates.  The data
agree with the predictions of the flat-$\Lambda$CDM model with
adiabatic initial perturbations. When considering the \acbar
data in combination with other contemporary CMB results, the
addition of a single parameter, $\Omega_\Lambda$, significantly
improves the fit to the data with a $\Delta\chi^2=20$.  

We find that the addition of the high sensitivity \acbar data
the the current CMB data does not lead to significant reductions
of the $1\sigma$ uncertainties in the canonical cosmological
parameters.  However, when considered in the eigenmode basis
of the parameter likelihood space, the addition of the \acbar data results
in a significant reduction of the observationally acceptable 
parameter space; this
indicates that the lack of improved errors on the fundamental parameters
probably results from degeneracies between the fiducial parameters.
The very high-$\ell$ points of the \acbar power spectrum indicate 
a slight excess of 
power at 150 GHz that is consistent with a value of $\sigma_8$ on the 
upper-end of limits from LSS if the excess is due to SZ emission from
clusters and is consistent with CBI and BIMA at 30 GHz.  
However, the \acbar data alone do not place to a $2\sigma$ 
lower-limit on the detection of this excess power.

It is clear from the measured CMB power spectrum in Figure \ref{fig:aps} 
that the \acbar
and CBI data are consistent with each other as well as the underlying $\Lambda$CDM
cosmological model.  This is significant for a number of reasons.  First, these
experiments have pushed the $\ell$-space coverage of the CMB power spectrum to
a factor of $\sim{3}$ larger than previous experiments.  
The remarkable agreement of the data with
the predictions of a $\Lambda$CDM cosmology across such a large range of angular
scale adds substantial credibility to the underlying model by testing the physics of the
damping region which are complementary to the accoustic peaks.  The high-$\ell$ data
have provided an essential test that the $\Lambda$CDM model has passed
with flying colors.  Second, the \acbar
and CBI data were taken by completely different experimental techniques (quasi-total
power bolometers versus interferometric measurements) at significantly different
observing frequencies (150 GHz for \acbar versus 30 GHz for CBI).  The sensitivity
of the two experiments to systematic effects 
are quite different and the consistency of the data sets is reasuring.

After publication of the \acbar results in late 2002, the WMAP team
released their phenomenal first-year power spectrum and cosmological parameters
\citep{hinshaw03,spergel03}.  This power spectrum runs out of steam in the 
vicinity of the third doppler peak ($\ell\sim 800$) and 
the \acbar and CBI data sets were
used to ``extend'' the WMAP power spectrum (forming the WMAPext
data set).  The damping tail data helps break some of the degeneracies
at low-$\ell$ 
and also provides marginal
evidence for a ``running'' scalar index, $dn_s/d\ln{k} \neq 0$ \citep{spergel03}
that is bolstered by incorporation of galaxy redshift and Lyman-$\alpha$ survey results.  
The value of the scalar index and running can be used to constrain inflationary models;
already CMB data have been used to rule out inflationary potentials of the form
$V(\phi) \propto \phi^4$ at $3\sigma$ \citep{kinney03}.  More precise
data in the damping tail region should assist in this endeavor.

The \acbar program has been primarily supported by NSF office of polar
programs grants OPP-8920223 and OPP-0091840.
This research used resources of the National Energy Research Scientific 
Computing Center, which is supported by 
the Office of Science of the U.S. Department of Energy under Contract 
No. DE-AC03-76SF00098. 
Chao-Lin Kuo acknowledges support from a Dr. and Mrs. CY Soong fellowship
and Marcus Runyan acknowledges support from a NASA Graduate Student 
Researchers Program fellowship.
Chris Cantalupo, Matthew Newcomb 
and Jeff Peterson acknowledge partial financial support 
from NASA LTSA grant NAG5-7926.

\tiny

\bibliographystyle{apj}
\bibliography{merged}

\begin{thebibliography}{25}
\expandafter\ifx\csname natexlab\endcsname\relax\def\natexlab#1{#1}\fi

\bibitem[{{Bennett} {et~al.}(1996){Bennett}, {Banday}, {Gorski}, {Hinshaw},
  {Jackson}, {Keegstra}, {Kogut}, {Smoot}, {Wilkinson}, \&
  {Wright}}]{bennett96}
{Bennett}, C.~L., {Banday}, A.~J., {Gorski}, K.~M., {Hinshaw}, G., {Jackson},
  P., {Keegstra}, P., {Kogut}, A., {Smoot}, G.~F., {Wilkinson}, D.~T., \&
  {Wright}, E.~L. 1996, \apjl, 464, L1

\bibitem[{{Bond} {et~al.}(1998){Bond}, {Jaffe}, \& {Knox}}]{bond98}
{Bond}, J.~R., {Jaffe}, A.~H., \& {Knox}, L. 1998, \prd, 57, 2117

\bibitem[{{Bond} {et~al.}(2002){Bond}, {Ruetalo}, {Wadsley}, \&
  {Gladders}}]{bond02b}
{Bond}, J.~R., {Ruetalo}, M.~I., {Wadsley}, J.~W., \& {Gladders}, M.~D. 2002,
  in ASP Conf. Ser. 257: AMiBA 2001: High-Z Clusters, Missing Baryons, and CMB
  Polarization, 15--+

\bibitem[{{Dawson} {et~al.}(2002){Dawson}, {Holzapfel}, {Carlstrom}, {Joy},
  {LaRoque}, \& {Reese}}]{dawson02}
{Dawson}, K.~S., {Holzapfel}, W.~L., {Carlstrom}, J.~E., {Joy}, M., {LaRoque},
  S.~J., \& {Reese}, E.~D. 2002, \apj, in press, astro-ph/020601

\bibitem[{{Efstathiou} \& {Bond}(1999)}]{efstathiou99}
{Efstathiou}, G. \& {Bond}, J.~R. 1999, \mnras, 304, 75

\bibitem[{{Finkbeiner} {et~al.}(1999){Finkbeiner}, {Davis}, \&
  {Schlegel}}]{finkbeiner99}
{Finkbeiner}, D.~P., {Davis}, M., \& {Schlegel}, D.~J. 1999, \apj, 524, 867

\bibitem[{Goldstein {et~al.}(2002)Goldstein, Ade, Bock, Bond, Cantalupo,
  Cantaldi, Daub, Holzapfel, Kuo, Lange, Lueker, Newcomb, Peterson, Pogosyan,
  Ruhl, Runyan, \& Torbet}]{goldstein02}
Goldstein, J., Ade, A.~R., Bock, J.~J., Bond, J.~R., Cantalupo, C., Cantaldi,
  C., Daub, M.~D., Holzapfel, W.~L., Kuo, C.~L., Lange, A.~E., Lueker, M.,
  Newcomb, M., Peterson, J.~B., Pogosyan, D., Ruhl, J., Runyan, M.~C., \&
  Torbet, E. 2002, ApJ submitted, astro-ph/0212517

\bibitem[{Hinshaw {et~al.}(2003)Hinshaw, Spergel, Verde, Hill, Meyer, Barnes,
  Bennett, Halpern, Jarosik, Kogut, Komatsu, Limon, Page, Tucker, Weiland,
  Wollack, \& Wright}]{hinshaw03}
Hinshaw, G., Spergel, D.~N., Verde, L., Hill, R.~S., Meyer, S.~S., Barnes, C.,
  Bennett, C.~L., Halpern, M., Jarosik, N., Kogut, A., Komatsu, E., Limon, M.,
  Page, L., Tucker, G.~S., Weiland, J., Wollack, E., \& Wright, E.~L. 2003,
  Submitted to ApJ, preprint: astro

\bibitem[{Kinney {et~al.}(2003)Kinney, Kolb, Melchiorri, \& Riotto}]{kinney03}
Kinney, W.~H., Kolb, E.~W., Melchiorri, A., \& Riotto, A. 2003,
  FERMILAB-Pub-03/117-A, preprint: hep-ph/0305130

\bibitem[{Kogut {et~al.}(2003)Kogut, Spergel, Barnes, Bennett, Halpern,
  Hinshaw, Jarosik, Limon, Meyer, Page, Tucker, Wollack, \& Wright}]{kogut03}
Kogut, A., Spergel, D.~N., Barnes, C., Bennett, C.~L., Halpern, M., Hinshaw,
  G., Jarosik, N., Limon, M., Meyer, S.~S., Page, L., Tucker, G., Wollack, E.,
  \& Wright, E.~L. 2003, Submitted to ApJ, astro

\bibitem[{{Komatsu} \& {Seljak}(2002)}]{komatsu02}
{Komatsu}, E. \& {Seljak}, U. 2002, \mnras, 336, 1256

\bibitem[{Kuo {et~al.}(2002)Kuo, Ade, Bock, Cantalupo, Daub, Goldstein,
  Holzapfel, Lange, Lueker, Newcomb, Peterson, Ruhl, Runyan, \& Torbet}]{kuo02}
Kuo, C.~L., Ade, P.~A.~R., Bock, J.~J., Cantalupo, C., Daub, M.~D., Goldstein,
  J.~H., Holzapfel, W.~L., Lange, A.~E., Lueker, M., Newcomb, M., Peterson,
  J.~B., Ruhl, J., Runyan, M.~C., \& Torbet, E. 2002, ApJ submitted,
  astro-ph/0212289

\bibitem[{{Lay} \& {Halverson}(2000)}]{lay00}
{Lay}, O.~P. \& {Halverson}, N.~W. 2000, \apj, 543, 787

\bibitem[{{Lee} {et~al.}(2001){Lee}, {Ade}, {Balbi}, {Bock}, {Borrill},
  {Boscaleri}, {de Bernardis}, {Ferreira}, {Hanany}, {Hristov}, {Jaffe},
  {Mauskopf}, {Netterfield}, {Pascale}, {Rabii}, {Richards}, {Smoot},
  {Stompor}, {Winant}, \& {Wu}}]{lee01}
{Lee}, A.~T., {Ade}, P., {Balbi}, A., {Bock}, J., {Borrill}, J., {Boscaleri},
  A., {de Bernardis}, P., {Ferreira}, P.~G., {Hanany}, S., {Hristov}, V.~V.,
  {Jaffe}, A.~H., {Mauskopf}, P.~D., {Netterfield}, C.~B., {Pascale}, E.,
  {Rabii}, B., {Richards}, P.~L., {Smoot}, G.~F., {Stompor}, R., {Winant},
  C.~D., \& {Wu}, J.~H.~P. 2001, \apjl, 561, L1

\bibitem[{Mason {et~al.}(2002)Mason, Pearson, Readhead, Shephard, Sievers,
  Udomprasert, Cartwright, Farmer, Padin, Myers, Bond, Contaldi, Pen, Prunet,
  Pogosyan, Carlstom, Kovac, Leitch, Pryke, Halverson, Holzapfel, Altamirano,
  Brofman, Casassus, May, \& Joy}]{mason02}
Mason, B.~S., Pearson, T., Readhead, A. C.~S., Shephard, M.~C., Sievers, J.~L.,
  Udomprasert, P.~S., Cartwright, J.~K., Farmer, A.~J., Padin, S., Myers,
  S.~T., Bond, J.~R., Contaldi, C.~R., Pen, U.-L., Prunet, S., Pogosyan, D.,
  Carlstom, J.~E., Kovac, J., Leitch, E., Pryke, C., Halverson, N., Holzapfel,
  W., Altamirano, P., Brofman, L., Casassus, S., May, J., \& Joy, M. 2002,
  Submitted to \apj, astro-ph/0205384

\bibitem[{{Netterfield} {et~al.}(2002){Netterfield}, {Ade}, {Bock}, {Bond},
  {Borrill}, {Boscaleri}, {Coble}, {Contaldi}, {Crill}, {de Bernardis},
  {Farese}, {Ganga}, {Giacometti}, {Hivon}, {Hristov}, {Iacoangeli}, {Jaffe},
  {Jones}, {Lange}, {Martinis}, {Masi}, {Mason}, {Mauskopf}, {Melchiorri},
  {Montroy}, {Pascale}, {Piacentini}, {Pogosyan}, {Pongetti}, {Prunet},
  {Romeo}, {Ruhl}, \& {Scaramuzzi}}]{netterfield02}
{Netterfield}, C.~B., {Ade}, P.~A.~R., {Bock}, J.~J., {Bond}, J.~R., {Borrill},
  J., {Boscaleri}, A., {Coble}, K., {Contaldi}, C.~R., {Crill}, B.~P., {de
  Bernardis}, P., {Farese}, P., {Ganga}, K., {Giacometti}, M., {Hivon}, E.,
  {Hristov}, V.~V., {Iacoangeli}, A., {Jaffe}, A.~H., {Jones}, W.~C., {Lange},
  A.~E., {Martinis}, L., {Masi}, S., {Mason}, P., {Mauskopf}, P.~D.,
  {Melchiorri}, A., {Montroy}, T., {Pascale}, E., {Piacentini}, F., {Pogosyan},
  D., {Pongetti}, F., {Prunet}, S., {Romeo}, G., {Ruhl}, J.~E., \&
  {Scaramuzzi}, F. 2002, \apj, 571, 604

\bibitem[{Oh {et~al.}(2003)Oh, Cooray, \& Kamionkowski}]{oh03}
Oh, S.~P., Cooray, A., \& Kamionkowski, M. 2003, preprint: astro-ph/0303007

\bibitem[{Peterson {et~al.}(2002)Peterson, Radford, Ade, Chamberlin, O'Kelly,
  Peterson, \& Schartman}]{peterson02}
Peterson, J.~B., Radford, S. J.~E., Ade, P. A.~R., Chamberlin, R.~A., O'Kelly,
  M.~J., Peterson, K.~M., \& Schartman, E. 2002, PASP submitted,
  astro-ph/0211134

\bibitem[{Romer {et~al.}(2003)Romer, Gomez, \& Collaboration}]{romer03}
Romer, A.~K., Gomez, P.~L., \& Collaboration, T.~A. 2003, in Carnegie
  Observatories Astrophysics Series, Vol. 3: Clusters of Galaxies: Probes of
  Cosmological Structure and Galaxy Evolution, ed. J.~S. Mulchaey, A.~Dressler,
  \& A.~Oemler (Pasadena: Carnegie Observatories),
  http://www.ociw.edu/ociw/symposia/series/symposium3/proceedings.html

\bibitem[{Runyan {et~al.}(2003{\natexlab{a}})Runyan, Ade, Bhatia, Bock, Daub,
  Goldstein, Haynes, Holzapfel, Kuo, Lange, Leong, Lueker, Newcomb, Peterson,
  Ruhl, Sirbi, Torbet, Tucker, Turner, \& Woolsey}]{runyan03a}
Runyan, M.~C., Ade, P.~A.~R., Bhatia, R.~S., Bock, J.~J., Daub, M.~D.,
  Goldstein, J.~H., Haynes, C.~V., Holzapfel, W.~L., Kuo, C.~L., Lange, A.~E.,
  Leong, J., Lueker, M., Newcomb, M., Peterson, J.~B., Ruhl, J., Sirbi, G.~I.,
  Torbet, E., Tucker, C., Turner, A.~D., \& Woolsey, D. 2003{\natexlab{a}}, ApJ
  submitted, astro-ph/0303515

\bibitem[{Runyan {et~al.}(2003{\natexlab{b}})Runyan, Ade, Bock, Cantalupo,
  Daub, Goldstein, Gomez, Holzapfel, Kuo, Lange, Lueker, Newcomb, Peterson,
  Romer, Ruhl, \& Torbet}]{runyan03b}
Runyan, M.~C., Ade, P.~A.~R., Bock, J.~J., Cantalupo, C., Daub, M.~D.,
  Goldstein, J.~H., Gomez, P., Holzapfel, W.~L., Kuo, C.~L., Lange, A.~E.,
  Lueker, M., Newcomb, M., Peterson, J.~B., Romer, A.~K., Ruhl, J., \& Torbet,
  E. 2003{\natexlab{b}}, In Preparation

\bibitem[{Spergel {et~al.}(2003)Spergel, Verde, Peiris, Komatsu, Nolta,
  Bennett, Halpern, Hinshaw, Jarosik, Kogut, Limon, Meyer, Page, Tucker,
  Weiland, Wollack, \& Wright}]{spergel03}
Spergel, D.~N., Verde, L., Peiris, H.~V., Komatsu, E., Nolta, M.~R., Bennett,
  C.~L., Halpern, M., Hinshaw, G., Jarosik, N., Kogut, A., Limon, M., Meyer,
  S.~S., Page, L., Tucker, G.~S., Weiland, J.~L., Wollack, E., \& Wright, E.~L.
  2003, Submitted to ApJ, Preprint: astro

\bibitem[{{Turner} {et~al.}(2001){Turner}, {Bock}, {Beeman}, {Glenn},
  {Hargrave}, {Hristov}, {Nguyen}, {Rahman}, {Sethuraman}, \&
  {Woodcraft}}]{turnerbolo}
{Turner}, A.~D., {Bock}, J.~J., {Beeman}, J.~W., {Glenn}, J., {Hargrave},
  P.~C., {Hristov}, V.~V., {Nguyen}, H.~T., {Rahman}, F., {Sethuraman}, S., \&
  {Woodcraft}, A.~L. 2001, \ao, 40, 4921

\bibitem[{{White}(2001)}]{white01}
{White}, M. 2001, \apj, 555, 88

\bibitem[{{Zhang} {et~al.}(2002){Zhang}, {Pen}, \& {Wang}}]{zhang02}
{Zhang}, P., {Pen}, U., \& {Wang}, B. 2002, \apj, 577, 555

\end{thebibliography}

\end{document}